\hfill {\bf UQMATH-arc-9704}
\baselineskip=16pt
\font\twelvebf=cmbx12

\vskip 15mm
\def\r{\rangle}
\def\l{\langle}
\def\[{[\![}
\def\]{]\!]}
\def\glm{{gl(m/\infty )}}

\def\Z_+={{\bf Z}_+}
\def\Z{{\bf Z}}
\def\C{{\bf C}}
\def\N{{\bf N}}

\leftskip 50pt

\noindent
{\twelvebf  Eigenvalues of Casimir operators for $gl(m/\infty )$ }

\vskip 1cm
\leftskip 50pt

\noindent
M.D. Gould and N.I. Stoilova\footnote{ $^{a)}$}{Permanent address: 
Institute for Nuclear
Research and
Nuclear Energy, 1784 Sofia, Bulgaria; Electronic mail: 
stoilova@inrne.acad.bg}

\noindent
Department of Mathematics, University of Queensland, Brisbane Qld
4072,
Australia

\vskip 48pt

\noindent
A full set of  Casimir operators for the Lie superalgebra
$\glm$ is constructed and shown to be well defined in the category
$O_{FS}$ generated by the highest weight irreducible
representations  with only a finite number of non-zero weight components.
The eigenvalues of these Casimir operators are determined
explicitly in terms of the highest weight. Characteristic identities
satisfied by certain (infinite) matrices with entries from $\glm$
are also determined.

\leftskip 0pt
\vskip 32pt
\noindent

\vfill\eject

\leftskip 0pt

\vskip 32pt 
\noindent
{\bf I. INTRODUCTION}

\bigskip 
During the last years the infinite dimensional Lie algebras and Lie
superalgebras play an important role in several areas of theoretical and
mathematical physics$^{1-9}.$ They have applications in the theory of
integrable field equations, string theory, two-dimensional statistical
models.  In addition these algebras are of interest as examples of Kac-Moody
Lie (super-)algebras of infinite type.

However, for these algebras  such a
fundamental concept as Casimir invariants has not yet been 
determined. The present paper is a step in solving this problem.

We construct a full set of Casimir operators for the infinite
dimensional general linear Lie superalgebra $\glm$  
corresponding to the natural
matrix realization, namely

$$\glm=\{ X=\left(\matrix{A&B\cr
C&D\cr}\right)|A\in M_{m\times m}, B\in M_{m\times \infty},
C\in M_{\infty \times m}, D\in M_{\infty \times \infty },\;$$
$$ \quad\quad all \; but \; a \;
finite \;
number
\; of \;X_{ij} \in \C \; are \; zero \}, \eqno (1)$$
where $M_{p\times q}$ is the space of all $p\times q$ complex matrices.
The even subalgebra $\glm _{\bar 0}$ 
has $B=0$ and $C=0$; 
the odd subspace $\glm_{\bar 1}$ 
has $A=0$ and $D=0$. 

A basis for the Lie superalgebra $\glm$ 
is given by the Weyl 
generators $E_{ij},\; i,j=-m+1,-m+2,\ldots , 0,1,\ldots .$ 
Assign to each index $i$ a degree $\langle i\rangle ,$ which is zero for 
$i\in -\Z_+$ and $1$ for $i\in \N$ (see the notation at the end of the
Introduction). Then the generator $E_{ij}$   is even (resp.
odd), if $\langle i\rangle +\langle j\rangle $ is an even 
(resp. odd) number. The multiplication 
($\equiv $the supercommutator) $\[\;,\;\]$ of $\glm$ 
is given by  the linear extension of the relations: 

$$
\[E_{ij},E_{kl}\]=\delta _{jk}E_{il}-
(-1)^{(\langle i\rangle +\langle j\rangle )(\langle k\rangle 
+\langle l\rangle )}\delta
_{il}E_{kj}. \eqno (2)
$$ 

We will consider the category 
$O_{FS}$ 
generated by all highest weight irreducible $\glm$ 
modules $V(\Lambda )$ 
with a finite number of non-zero highest weight components $\Lambda _i$
of the highest weight
$$\Lambda  \equiv (\Lambda _{-m+1}, \Lambda _{-m+2}, \ldots , \Lambda _0,
\Lambda _1,\ldots , \Lambda _k, 0,
 0,\ldots )\equiv
(\Lambda _{-m+1}, \Lambda _{-m+2}, \ldots , \Lambda _0,
\Lambda _1,\ldots , \Lambda _k, \dot{0}).
 \eqno (3)$$
The highest weight $\Lambda$ 
of $V(\Lambda )$ 
uniquely characterized the module and 
satisfies the conditions:
$$
\Lambda _i-\Lambda _{i+1} \in \Z_+ ,\quad 
 \forall i\neq 0. \eqno (4)
$$

Denote by $H$ the Cartan subalgebra of $\glm.$ The dual 
space $H^*$ of $H$ is described by the forms 
$\varepsilon _i,\;i=-m+1, -m+1, \ldots ,$ 
where $\varepsilon _i:X\rightarrow A_{ii},$ for $-m+1\leq i\leq 0$
and $\varepsilon _i:X\rightarrow D_{ii}, \;\;\forall i\in \N,$
and $X$ is given  by (1) only for diagonal $X$. On $H^*$ 
there is a bilinear form $ (\;,\;)$ 
 defined by 

$$
\eqalignno{
& (\epsilon _i, \epsilon _j)=\delta _{ij}, \;\;\quad  for \;\; 
-m+1\leq i,j\leq 0; &\cr
&(\epsilon _i, \epsilon _j)=0, \;\;\;\;\quad for \;\; 
-m+1\leq i\leq 0\;\; and \;\;j\in \N ;& (5) \cr
&(\epsilon _i, \epsilon _j)=-\delta _{ij}, \;\; for \;\; \;\;
i,j\in \N .&  \cr
}
$$

\noindent
 The roots $\varepsilon _i-\varepsilon_j \;(i\neq j)$
 of $\glm$ are the non-zero weights of the adjoint
representation. The positive roots are those given by the set:
$$\Phi ^+= \{\varepsilon _i-\varepsilon _j|
i<j, \; i,j=-m+1,-m+2,\ldots  \}. \eqno (6)$$

\noindent
Define
$$\rho ={1\over 2}\sum_{i=-m+1}^{0}(1-2i-2m)\epsilon _i+
{1\over 2}\sum_{i=1}^{\infty }(1-2i+2m)\epsilon _i. \eqno (7)$$
Let  $D_n$ be the set of $\glm$ weights: 
$$D_n=\{ \nu |\nu =(\nu_{-m+1},\ldots , \nu_0,\nu_1,\ldots ,
\nu_n, \dot{0}), \;\; 
\nu_i \in \Z_+,\;\; i=-m+1,-m+2,\ldots ,n-1,\;\; \nu_n\in \N \} , \eqno(8)$$
and let $D_n^+\subset D_n $ be the subset of integral dominant 
weights in $D_n:$
$$D_n^+=\{ \nu |\nu \in D_n,\; 
\nu _i-\nu _{i+1}\in \Z_+, \;\; \forall i\neq 0\} .
 \eqno (9)$$

\noindent
Note that 
if $\nu $ is a weight in $V(\Lambda ),\; \Lambda \in D_k^+,$ 
then $\nu \in D_n,$ for some 
$n\in \Z_+.$ 

In Section II we construct a full set of Casimir operators 
convergent on each module $V(\Lambda )$. The eigenvalues of
these Casimir invariants for all modules from the category 
$O_{FS}$ are computed in Section III.
In Section IV  we present a derivation of the polynomial 
identities satisfied by
certain matrices with entries from $\glm$.

Throughout the paper we use the following notation:

\bigskip
irrep(s) - irreducible representation(s);

$\C$ - the complex numbers;

$\Z_+$ - all non-negative integers;

$\N$ - all positive integers;

$U(A)$ - the universal enveloping algebra of $A;$

\smallskip

$
\l i\r =\cases { 0 & for $i\in -\Z_+$ \cr
1 & for $i\in \N$. \cr}
$
\vskip 1cm
\noindent
{\bf II. CONSTRUCTION OF CASIMIR OPERATORS }

\smallskip
An obvious invariant for $\glm$ 
is the first order 
invariant

$$I_1=\sum_{i=-m+1}^\infty E_{ii}. \eqno(10)$$
It is not clear, however, how to 
construct appropriate higher order Casimir operators for 
$\glm$. 
Let us first consider
the second order invariant $I_2^{(m,n)}$ of $gl(m/n):$

$$
\eqalignno{
& I_2^{(m,n)}=\sum_{i,j=-m+1}^n(-1)^{\l j \r }E_{ij}E_{ji}=
\sum_{i,j=-m+1}^0E_{ij}E_{ji}- \sum_{i,j=1}^nE_{ij}E_{ji}+
\sum_{i=1}^n\;\;\sum_{j=-m+1}^0E_{ij}E_{ji} &\cr
&&\cr
& - \sum_{i=-m+1}^0\;\;\sum_{j=1}^nE_{ij}E_{ji}= 
\sum_{i=-m+1}^0\;\;\sum_{j<i=-m+1}^0
E_{ij}E_{ji}+\sum_{i=-m+1}^0\;\;\sum_{j>i=-m+1}^0
E_{ij}E_{ji}+\sum_{i=-m+1}^0 E_{ii}^2 &\cr
&&\cr
& -\sum_{i=1}^n\sum_{j<i=1}^n
E_{ij}E_{ji}-\sum_{i=1}^n\sum_{j>i=1}^n
E_{ij}E_{ji}-\sum_{i=1}^n E_{ii}^2+ 
 2\sum_{i=1}^n\;\;\sum_{j=-m+1}^0
E_{ij}E_{ji}-\sum_{i=-m+1}^0\;\;\sum_{j=1}^n
(E_{ii}+E_{jj}) & \cr  
&&\cr
& =2\sum_{i=-m+1}^0\;\;\sum_{j<i=-m+1}^0
E_{ij}E_{ji}+\sum_{i=-m+1}^0\;\;\sum_{j>i=-m+1}^0
(E_{ii}-E_{jj})
+\sum_{i=-m+1}^0 E_{ii}^2 -
2\sum_{i=1}^n\sum_{j<i=1}^n
E_{ij}E_{ji}&\cr
&&\cr
&-\sum_{i=1}^n\sum_{j>i=1}^n
(E_{ii}-E_{jj})
-\sum_{i=1}^n E_{ii}^2 + 
 2\sum_{i=1}^n\;\;\sum_{j=-m+1}^0
E_{ij}E_{ji}-n\sum_{i=-m+1}^0 
E_{ii}-m\sum_{i=1}^n E_{ii}&\cr
&&\cr
&=2\sum_{i=-m+1}^n\;\;\sum_{j<i=-m+1}^n(-1)^{\l j\r}
E_{ij}E_{ji}+\sum_{i=-m+1}^0 E_{ii}(E_{ii}+1-m-2i)-
\sum_{i=1}^n E_{ii}(E_{ii}+1+n-2i)&\cr
&&\cr
&-n\sum_{i=-m+1}^0E_{ii} -m\sum_{i=1}^nE_{ii} &\cr
&&\cr
&=2\sum_{i=-m+1}^n\;\;\sum_{j<i=-m+1}^n(-1)^{\l j\r}
E_{ij}E_{ji}+\sum_{i=-m+1}^n (-1)^{\l i\r}
E_{ii}(E_{ii}+1-2i)-
(m+n)I_1^{(m,n)}&\cr
& &\cr
& =2\sum_{i=-m+1}^n\;\;\sum_{j<i=-m+1}^n(-1)^{\l j\r}
E_{ij}E_{ji}+\sum_{i=-m+1}^n (-1)^{\l i\r}
E_{ii}(E_{ii}+1-2i)-
2mI_1^{(m,n)}+ (m-n)I_1^{(m,n)},&\cr
&& (11) \cr
}
$$
where $I_1^{(m,n)}\equiv \sum_{i=-m+1}^nE_{ii}$ is the first order 
invariant of
$gl(m/n).$ Due to the last term in (11) the $gl(m/n)$ second order invariant
diverges as $n\rightarrow \infty .$ Eliminating the last term in (11) 
(the rest 
of the expression is also an invariant) and taking the limit 
$n\rightarrow \infty $ one obtains the following quadratic Casimir for 
$\glm$:

$$
I_2=2\sum_{i=-m+1}^\infty\;\;\sum_{j<i=-m+1}^\infty
(-1)^{\l j\r}
E_{ij}E_{ji}+\sum_{i=-m+1}^\infty (-1)^{\l i\r}
E_{ii}(E_{ii}+1-2i)-
2mI_1, \eqno (12)
$$
which is convergent (see formula (21)) on the category $O_{FS}$ 
of irreps considered. On 
$V(\Lambda ), \;\Lambda \in D_k^+,\;\; I_2$ 
takes constant value
$$\chi _{\Lambda } (I_2)=\sum_{i=-m+1}^k \left((-1)^{\l i\r }\Lambda _i 
(\Lambda _i+1-2i)-2m\Lambda _i\right)
=(\Lambda, \Lambda+2\rho). \eqno (13)$$
This consideration shows  how to construct  the higher order
Casimir operators of $\glm.$

Introduce to this end the characteristic matrix
$$A_i^{\;j}=(-1)^{\l i\r \l j\r }E_{ji}. \eqno (14)$$
Define the powers of the matrix $A$ 
recursively by
$$\left( A^q \right) _i^{\;j}=\sum_{k=-m+1}^{\infty }A_i^{\;k}
(A^{q-1})_k^{\;j}, \quad\quad 
[(A^0)_i^{\;j}\equiv \delta _{ij}]. \eqno (15)$$

 Using  induction and the $\glm$
commutation relations (2) one obtains:

{\it Proposition 1:} 

$$\[E_{kl}, (A^q)_i^{\;j}\]=(-1)^{(\l k\r +\l l\r)\l i\r}
\left( \delta _{lj}(A^q)_i^{\;k}-\delta _{ik}
(A^q)_l^{\;j}\right). \eqno (16)$$
\hskip 16 cm $[]$

\noindent
Therefore the matrix supertraces
$$str(A^q)\equiv \sum_{i=-m+1}^{\infty}(-1)^{\l i\r }
(A^q)_i^{\;i} \eqno (17)$$
are formally Casimir operators. They are, however, divergent except for 
$q=1$ in
which case we obtain the first order invariant (10). Our purpose is 
to construct a full set of Casimir invariants which
are well defined and convergent on the category $O_{FS}.$ 

{\bf Theorem 1:} {\it The Casimir operators defined recursively by}
$$
\eqalignno{
& I_1=\sum_{i=-m+1}^{\infty} 
(-1)^{\l i\r }A_i^{\;i}=str(A); &  \cr
& I_q=\sum_{i=-m+1}^{\infty} 
(-1)^{\l i\r }\left [(A^q)_i^{\;i}
-I_{q-1}\right]=
str\left[ A^q-I_{q-1}\right] & (18) \cr
}
$$
{\it form a full set of convergent $\glm$ Casimir operators 
on each module }
$V(\Lambda )\in O_{FS}$.\hskip 3cm $[]$

Observe that the operators $I_q$  are indeed Casimir invariants
(see {\it Proposition 1}). Then it remains to prove they are 
convergent on the category
$O_{FS}.$ We will do this by induction. Consider 
first the case $q=2:$

$$
\eqalignno{
& I_2\equiv \sum_{j=-m+1}^{\infty }(-1)^{\l j\r }
\left[(A^2)_j^{\;j}-I_1\right]=
\sum_{j=-m+1}^{0 }\left[ \sum_{i=-m+1}^{\infty }E_{ij}E_{ji}-
I_1\right]&\cr
&&\cr
&- \sum_{j=1}^{\infty }\left[ \sum_{i=-m+1}^{\infty
}E_{ij}E_{ji}-I_1\right] 
=\sum_{j=-m+1}^{0}\;\;\sum_{i=-m+1}^{0 }
E_{ij}E_{ji}+
\sum_{j=-m+1}^{0}\;\sum_{i=1}^{\infty}
E_{ij}E_{ji}-mI_1&\cr
&&\cr
&-\sum_{j=1}^{\infty }\;\sum_{i=-m+1}^0
E_{ij}E_{ji}-
\sum_{j=1}^{\infty }\left[ \sum_{i=1}^\infty E_{ij}E_{ji}-
I_1\right]
 & \cr
&&\cr
& =2 \sum_{i=-m+1}^{0}\;\;\sum_{j<i=-m+1}^0
E_{ij}E_{ji}+\sum_{i=-m+1}^{0 }
E_{ii}(E_{ii}+1-m-2i)-mI_1+
2\sum_{j=-m+1}^{0} \;\sum_{i=1}^{\infty}
E_{ij}E_{ji}&\cr
&&\cr
& -\sum_{j=1}^{\infty }\;\;\sum_{i=-m+1}^0(E_{ii}+E_{jj})
 -\sum_{j=1}^{\infty }\left[ 2\sum_{i>j=1}^{\infty }
E_{ij}E_{ji}+
\sum_{i<j=1}^{\infty }
(E_{ii}-E_{jj})+E_{jj}^2-I_1\right]&\cr
&&\cr
& =2\sum_{i=-m+1}^{\infty }\;\;\sum_{j<i=-m+1}^{\infty }
(-1)^{\l j\r }E_{ij}E_{ji}+
\sum_{i=-m+1}^{0}
E_{ii}(E_{ii}+1-m-2i)-mI_1-
m\sum_{i=1}^\infty E_{ii}
&\cr
&&\cr
&-\sum_{j=1}^\infty\left[ \sum_{i<j=1}^\infty (E_{ii}-E_{jj})+
E_{jj}^2-\sum_{i=1}^\infty E_{ii}\right] & \cr
&&\cr
&=2\sum_{i=-m+1}^{\infty }\;\;\sum_{j<i=-m+1}^{\infty }
(-1)^{\l j\r }E_{ij}E_{ji}+
\sum_{i=-m+1}^{0}
E_{ii}(E_{ii}+1-2i)
-2mI_1-\sum_{i=1}^\infty E_{ii}(E_{ii}+1-2i), & \cr
&&  (19) \cr
}
$$
which agrees with the definition (12).

Now let ${ \it v}\in V(\Lambda ),\;\Lambda \in D_k^+,\;$ 
 be an arbitrary
weight  vector. Then the weight of $ {\it v}$ has the form
$$\nu =(\nu _{-m+1},\nu _{-m+2}, \ldots , \nu _{0}, 
\ldots , \nu _{r}, \dot{0}). \eqno (20)$$
Since
$$A_i^{\;j}{\it v}=(-1)^{\l i\r \l j\r}E_{ji}{\it v}=0, 
\;\; \; \forall i>r, \eqno (21)$$
the second order invariant $I_2$ is convergent on each
$V(\Lambda )\in O_{FS} $ [c.f. formula (13)].

Applying {\it Proposition 1} and (21), for $i>r$ one  obtains
$$
\eqalignno{
& (A^q)_i^{\;i}{\it v}=\sum_{j=-m+1}^{\infty }
A_i^{\;j}(A^{q-1})_j^{\;i}{\it v}
=\sum_{j=-m+1}^{\infty }(-1)^{\l i\r \l j \r}E_{ji}
(A^{q-1})_j^{\;i}{\it v} & \cr
& =
\sum_{j=-m+1}^{\infty }(-1)^{\l i\r \l j\r }
\left\{ (-1)^{(\l j\r +\l i\r )\l j\r }\left[ 
(A^{q-1})_j^{\;j}-(A^{q-1})_i^{\;i}\right]
 {\it v}+(-1)^{(\l i\r +\l j\r )}
(A^{q-1})_j^{\;i}E_{ji}{\it v} \right\} &\cr
&
 =\sum_{j=-m+1}^{\infty }(-1)^{\l j\r }\left[ (A^{q-1})_j^{\;j}
-(A^{q-1})_i^{\;i}\right] {\it v}.
& (22) \cr
}
$$
For the case $q=2$ we have

$$(A^2)_i^{\;i}{\it v}=\sum_{j=-m+1}^{\infty }(-1)^{\l j\r }\left[ 
A_j^{\;j}-A_i^{\;i}\right] {\it v}=\sum_{j=-m+1}^{\infty }E_{jj}{\it v}
=I_1{\it v}, \;\; 
\forall i>r \eqno (23)$$
so that
$$\left( (A^2)_i^{\;i}-I_1\right) {\it v}=0,\; \forall i>r, 
\eqno (24)$$
which is another proof for the convergence of $I_2.$
More generally

{\it Proposition 2:} {\it For any weight vector} ${ \it v}\in V(\Lambda ),$ 
{\it and $q\in \N$
there exist 
$r\in \N$ such that}
$$\left((A^q)_i^{\;i}-I_{q-1}\right)
{\it v}=0,\;\;\forall i>r. \eqno(25)$$

{\it Proof:} We proceed by induction. Assume $v$ has weight $\nu $ as in
(20). Formula (25) is valid for $q=2$ (24). 
Let the result be
true for a given $q$, i.e.
$$(A^q)_i^{\;i} {\it v}=I_{q-1}{\it v},\;\;\forall i>r.$$
Then (see (22))
$$(A^{q+1})_i^{\;i}{\it v}=
\sum_{j=-m+1}^{\infty }(-1)^{\l j\r }\left[ 
(A^{q})_j^{\;j}-(A^{q})_i^{\;i}\right] {\it v}=
\sum_{j=-m+1}^{\infty }(-1)^{\l j\r}
\left[ (A^{q})_j^{\;j}-I_{q-1}
\right] {\it v}=I_q{\it v},
\quad \forall i>r, \eqno (26)$$
which proves (25). 
\hskip 13 cm $[]$

\bigskip

$I_q$ (18) is convergent on each $V(\Lambda )$ for $q=2.$ Assume it is
well defined and convergent on $ V(\Lambda )$ 
for a given
$q.$ Then, with $v$ as in (25), we have
$$I_{q+1}{\it v}\equiv \sum_{i=-m+1}^{\infty}
(-1)^{\l i\r }\left[ (A^{q+1})_i^{\;i}-I_q
\right] {\it v}=\sum_{i=-m+1}^{r}(-1)^{\l i\r }
\left[ (A^{q+1})_i^{\;i}-I_q
\right] {\it v}
$$
$$
=\sum_{i=-m+1}^{r}(-1)^{\l i \r }
(A^{q+1})_i^{\;i}{\it v} +(r-m)I_q
 {\it v}. \eqno (27)$$
Therefore $I_{q+1}$ is convergent and well defined 
on $V(\Lambda ).$

This completes the (inductive) proof of {\it Theorem 1}.

\vskip 1cm
\noindent
{\bf III. EIGENVALUE FORMULA FOR CASIMIR OPERATORS}

\smallskip
In this section we apply our previous results to evaluate the spectrum of
the operators (18).

Let $v\in V(\Lambda ),$ be an arbitrary vector of weight 
$\nu =(\nu_{-m+1}, \nu_{-m+2} \ldots ,\nu_0, \nu_1,
\ldots , \nu _r, \dot{0}).$ Then, keeping in mind {\it
Proposition 1,} the fact that $(A^{q-1})_k^{\;j}$ has weight $\varepsilon
_j-\varepsilon _k$ under the adjoint representation of $\glm$ and
that all vectors of $V(\Lambda )$ have weight components 
$\nu _i$ in $\Z_+,$ we must
have for $j\leq r$
$$(A^{q-1})_k^{\;j}v=0, \quad \forall k>r. \eqno (28)$$
Therefore
$$(A^q)_i^{\;j}v=\sum_{k=-m+1}^{\infty }A_i^{\;k}(A^{q-1})_k^{\;j}v
=\sum_{k=-m+1}^{r }A_i^{\;k}(A^{q-1})_k^{\;j}v. \eqno (29)$$
Proceeding recursively  we may therefore write
$$(A^q)_i^{\;j}v=(\bar{A}^{q})_i^{\;j}v, \quad \forall i,j=-m+1,
-m+2, \ldots, r, \eqno (30)$$
where $(\bar{A})_i^{\;j}=(-1)^{\l i\r \l j\r }E_{ji}, \;\; 
\forall i,j=-m+1,\ldots , r,$ is the
$gl(m/r)$ characteristic matrix, and the powers of the matrix 
$\bar{A}$ are
defined by (15) with $i,j,k=-m+1,\ldots ,r$ and 
$\bar{A}$ instead of $A.$
It follows then that the formula (27) can be written as:
$$I_{q}{\it v}= \sum_{i=-m+1}^{r}(-1)^{\l i\r }
\left[ (\bar{A}^{q})_i^{\;i}-I_{q-1}
\right] {\it v}=\left[ I_q^{\;(m,r)}-(m-r)I_{q-1}
\right] {\it v}, \eqno (31)$$
with 
$$I_q^{\;(m,r)}=\sum_{i=-m+1}^r(-1)^{\l i\r }
(\bar{A}^q)_i^{\;i}, \eqno(32)$$
 being the 
$q^{th}$ order invariant of $gl(m/r).$
Formula (31) is valid $\forall q\in \N,$ which gives a recursion relation
for the $I_q$ with initial condition
$$I_1v=\chi _{\Lambda }(I_1)v. \eqno (33)$$
In particular it follows from (31) that the invariants 
$I_q$ are certainly
convergent on all weight vectors $v\in V(\Lambda ).$

To determine the eigenvalues of $I_q$ let $v=v_{\Lambda }^+$ be the highest
weight vector of the  $V(\Lambda )$ module
and let
$$\Lambda =(\bar{\Lambda }, \dot{0})\in D_k^+, \quad 
\bar{\Lambda }\equiv 
(\Lambda _{-m+1}, \Lambda _{-m+2},\ldots ,\Lambda _0,
\Lambda _1,\ldots , \Lambda _k). \eqno (34)$$
Then for the eigenvalues of the $I_q$ 
one obtains the recursion relation
(see (31)): 
$$\chi_{\Lambda }(I_q)=
\chi_{\bar{\Lambda }}(I_q^{\;(m,k)})-
(m-k)\chi _{\Lambda }(I_{q-1}), 
\quad \chi_{\Lambda }(I_1)=\sum_{i=-m+1}^k
\Lambda _i, \eqno (35)$$
where $\chi_{\bar{\Lambda }}(I_q^{\;(m,k)})$ is the eigenvalue of 
the $q^{th}$
order invariant (32) of $gl(m/k)$ on the irreducible $gl(m/k)$ 
module with
highest weight $\bar{\Lambda };$ the latter is given explicitly by$^{10}$
$$\chi_{\bar{\Lambda }}(I_q^{\;(m,k)})=\sum_{i=-m+1}^k (-1)^{\l i \r }
\alpha _i^q \prod_{j\neq i=-m+1}^k\left( {\alpha _i-\alpha _j+
(-1)^{\l j\r} \over{\alpha _i-\alpha _j}}\right), \eqno (36)$$
where
$$\alpha _i=(-1)^{\l i\r}\left( \Lambda _i-i+1\right)-m. $$
Therefore we  obtain for the eigenvalues of the Casimir operators 
$I_q$
$$\chi _{\Lambda }(I_q)=\sum_{i=-m+1}^k
(-1)^{\l i\r}P_q(\alpha _i)
\prod_{j\neq i=-m+1}^k \left( {\alpha _i-\alpha _j+
(-1)^{\l j\r} \over{\alpha _i-\alpha _j}}\right),
 \eqno (37)$$
for suitable polynomials $P_q(x)$ which, from Eq. (35), satisfy the recursion
relation
$$P_q(x)=x^q-(m-k)P_{q-1}(x), \quad P_1(x)=x. \eqno (38)$$
In particular
$$
\eqalignno{
& P_2(x)=x^2-(m-k)x=x{x^2-(m-k)^2\over{x+(m-k)}}; & (39a) \cr
& P_3(x)=x^3-(m-k)(x^2-(m-k)x)=x{x^3+(m-k)^3\over{x+(m-k)}}, & (39b) \cr
}
$$
and more generally, it is easily established by induction that
$$P_q(x)=x{x^q-(-1)^q(m-k)^q\over{x+(m-k)}}. \eqno (40)$$ 
Thus we have

{\bf Theorem 2:} {\it The eigenvalues of the Casimir operators
$I_q$ (18),
 on the
irreducible  $\glm$ module $V(\Lambda )$, 
$\Lambda \in D_k^+$
are given by }
$$\chi_{\Lambda }(I_q)
=\sum_{i=-m+1}^k (-1)^{\l i\r}\alpha _i\left({\alpha _i^q-
(-1)^{q}(m-k)^q\over{\alpha _i+(m-k)}}\right) 
\prod_{j\neq i=-m+1}^k \left( {\alpha _i-\alpha _j+
(-1)^{\l j\r} \over{\alpha _i-\alpha _j}}\right),
$$
$$
\quad
where \;\;
\alpha _i=(-1)^{\l i\r}\left( \Lambda _i-i+1\right)-m. 
\eqno (41)$$
\hskip 16cm $[]$

\vfill\eject
\noindent
{\bf IV. POLYNOMIAL IDENTITIES}

\smallskip
Let $\Delta $ be the comultiplication on the enveloping algebra
$U[\glm ]$ of $\glm $ ($\Delta (E_{ij})=E_{ij}\otimes 1+
1\otimes E_{ij}, \;\;i,j=-m+1,-m+2,\ldots $ with $1$ 
being the unit in  $U[\glm ]$).
Applying $\Delta $ to the second order Casimir operator (12) of
$\glm$ we obtain:
$$\Delta (I_2)=I_2\otimes 1+1\otimes I_2+2\sum_{i,j=-m+1}^{\infty }
(-1)^{\l j\r }E_{ij}\otimes E_{ji}. \eqno (42)$$
Therefore
$$\sum_{i,j=-m+1}^{\infty }(-1)^{\l j\r }E_{ij}\otimes E_{ji}=
{1\over 2}\left[\Delta (I_2)-I_2\otimes 1-1\otimes I_2\right]. 
\eqno (43) $$
Denote by $\pi_{\varepsilon _{-m+1}}$ the irrep of 
$\glm $ afforded by 
$V(\varepsilon _{-m+1}).$ The weight spectrum for the vector module 
$V(\varepsilon _{-m+1})$ consists of all weights $\varepsilon _i, \;
i=-m+1,-m+2, \ldots , $ each occurring exactly once. 
Denote by $e_{ij}, \; i,j=-m+1,-m+2,\ldots $
 the generators on this space
$$\pi_{\varepsilon _{-m+1}}(E_{ij})=e_{ij}, \eqno (44)$$
with $e_{ij}$ an elementary matrix.

Introduce the characteristic matrix
$$A=
{1\over 2}(\pi_{\varepsilon _{-m+1}}\otimes 1)\left[\Delta (I_2)-
I_2\otimes 1-1\otimes I_2\right]. \eqno (45)$$ 
Therefore 
$$
A_k^{\;\;l}=\sum_{i,j=-m+1}^\infty \; (-1)^{(\l i\r +\l j\r )\l l\r }
\pi_{\varepsilon _{-m+1}}(E_{ij})_{kl}(-1)^{\l j\r }E_{ji}=
(-1)^{\l k \r \l l \r}E_{lk}. \eqno(46)
$$

The matrix 
$A$ is the infinite matrix introduced in Sec. II (see (14))
and the entries  of the matrix powers $A^q$ are given recursively by (15). 
We will see that the characteristic matrix satisfies a polynomial identity
acting on the $\glm$ module $V(\Lambda ), \; \Lambda \in D_k^+.$ 
Let $\pi_{\Lambda }$ be the representation afforded by $V(\Lambda ).$
From Eq.
(45), acting on $V(\Lambda )$ we may interpret $A$ as an invariant
operator on the tensor product module 
$V(\varepsilon _{-m+1})\otimes V(\Lambda ):$
$$A\equiv {1\over 2}(\pi_{\varepsilon _{-m+1}}\otimes \pi_{\Lambda })
\left[ \Delta (I_2)-I_2\otimes 1-1\otimes I_2\right]. \eqno (47)$$
Following Ref. 11 it is easy to see that the tensor product 
space admits a filtration of submodules

$$V(\varepsilon _{-m+1})\otimes V(\Lambda )=
V_{k+1}\supseteq V_k\supseteq \ldots  V_0 \supseteq
\ldots  \supseteq V_{-m+1}  \supseteq (0),\eqno(48)$$ 
where each factor module  $M_i=V_i/V_{i+1},$ if non-zero,
is indecomposable and cyclically generated by a highest weight 
vector of weight  $\Lambda +\varepsilon _i.$ We emphasize that 
$M_i$ is only non-zero when $\Lambda +\varepsilon _i$ is 
integral dominant.  Then it follows that the generalized eigenvalues
of $A$ on the tensor product space are given by
$${1\over 2}\left[ \chi_{\Lambda +\varepsilon_i}(I_2)-
\chi_{\varepsilon_{-m+1}}(I_2)-\chi_{\Lambda }(I_2)\right]=
{1\over 2}\left[(\Lambda +\varepsilon _i, \Lambda +\varepsilon _i+2\rho )-
(\varepsilon _{-m+1}, \varepsilon _{-m+1}+2\rho )-
(\Lambda , \Lambda +2\rho )\right]$$
$$
=(-1)^{\l i\r }\left( \Lambda _i+1-i\right) -m, \eqno (49)$$  
(see {\it Theorem 2}). Thus we have

{\bf Theorem 3:} {\it On each $\glm$ module $V(\Lambda )$, 
$\Lambda \in D_k^+$ the
characteristic matrix satisfies the polynomial identity}
$$\prod_{i=-m+1}^{k+1}\left( A-\alpha _i\right)=0, 
 \eqno(50)$$
{\it with $\;\alpha _i=(-1)^{\l i\r }\left( 
\Lambda _i+1-i\right) -m\;$ the characteristic roots.}
\hskip 6.5cm $[]$

\noindent
Note that the characteristic identities (50) are the $\glm $ 
counterpart of
the polynomial identities encountered for $gl(m/n)$ by 
Jarvis and Green
$^{12}$ (more precisely their adjoint identities).
\vskip 1cm

\bigskip
\noindent
{\bf ACKNOWLEDGMENTS}

\smallskip
One of us (N.I.S.) is grateful  for the kind
invitation to work in the mathematical physics group at the 
Department of Mathematics
in University of Queensland.  The
work was supported by the Australian Research Council and by the
Grant $\Phi -416$ of the Bulgarian Foundation for Scientific
Research.

\vskip 12pt
{\settabs \+  $^{11}$& I. Patera, T. D. Palev, Theoretical
   interpretation of the experiments on the elastic \cr

\+ $^1$  & V.G. Kac, {\it Infinite Dimensional Lie Algebras} 
(Cambridge University Press, Cambridge, 1985), Vol. {\bf 44}. \cr

\+ $^2$ & V.G. Kac  and A.K. Raina, {\it Bombay Lectures on Highest Weight
         Representations of Infinite}  \cr
\+     &  {\it Dimensional Lie Algebras} in Advanced Series in
          Mathematics (World Scientific, Singapore, 1987), Vol. {\bf 2}. \cr
 
\+ $^3$ & E. Date, M. Jimbo, M. Kashiwara and T. Miwa, 
 Publ. RIMS Kyoto Univ. {\bf 18}, 1077 (1982). \cr

\+ $^4$ & M. Sato, RIMS Kokyoroku {\bf 439}, 30 (1981). \cr

\+ $^5$ & P. Goddard and D. Olive, Int. J. Mod. Phys. {\bf A 1}, 303
         (1986). \cr

\+ $^6$ & B. Feigin and D. Fuchs,  
    "Representations of the Virasoro algebra," in 
     {\it Representations of Infinite} \cr
\+     & {\it Dimensional Lie Groups and Lie Algebras} 
      (Gorgon and  Breach, New York, 1989). \cr

\+ $^7$  & V.G. Kac, J.W. van de Leur, Ann. Inst. Fourier, Grenoble 
 {\bf 37}, 99 (1987). \cr

\+ $^8$  & V.G. Kac, J.W. van de Leur, {\it Infinite Dimensional Lie
Algebras and Groups,} edited by V.G. Kac, \cr
\+       & Advanced Series in Mathematical 
Physics {\bf 7} (World Scientific, Singapore, 1989), pp. 369-406. \cr

\+ $^9$  & K. Ikeda, Lett. Math. Phys. {\bf 14}, 321 (1987).
 {\bf 37}, 99 (1987). \cr

\+ $^{10}$ & J.R. Links, R.B. Zhang, Journ. Math. Phys. {\bf 34},
6016 (1993); M.D. Gould, J.R. Links, Y.-Z. Zhang, \cr
\+         & Lett. Math. Phys. {\bf 36}, 415 (1996). \cr

\+ $^{11}$ & M.D. Gould, J. Austral. Math. Soc. Ser. B 
 {\bf 28}, 310 (1987). \cr

\+ $^{12}$ & P.D. Jarvis, H.S. Green, Journ. Math. Phys. {\bf 20},
2115 (1979). \cr

\end